\documentstyle{article}

\begin{document}

\title{\bf
Fractals in Linear Ordinary Differential \\
Equations}

\author{
{\bf
Dhurjati Prasad Datta}\\
\normalsize North Eastern Regional Institute of
Science and Technology\\
Itanagar-791101, Arunachal Pradesh, India \\ \\
email: dpd@nerist.ernet.in}

\baselineskip = 16pt

\date{2 May 1997}

\maketitle

\bigskip

\begin{abstract}
  We prove the existence of fractal solutions to a class of linear
ordinary differential equations.This reveals the possibility
of chaos in the very short time limit of the evolution even of a 
linear one  dimensional dynamical system.
\end{abstract}

\newpage

\section{Introduction}

	Understanding the complexities in a nonlinear dynamical system is of   
great interest in contemporary  sciences. Both chaos and fractals seem 
to offer two important ingredients towards this effort.
Even a simple quadratic nonlinearity as represented by the logistic map
is known to generate, for a sufficiently large control parameter ,
a fractal attractor, indicating the onset of the deterministic chaos
in the model. A linear equation, on the other hand ,is supposed to be
immune to fractal attractors (solutions). In this paper
we,however, report, on the contrary, a rather surprising fact: the 
simplest linear ODE  of the form $dy/dt = h(t)y$ admits fractal 
solutions, besides the standard exponential solution. This observation 
follows directly from our recent results indicating a deep relationship
between the quantal geometric phase and the concept of time derived
intrinsically from the (quantal) evolution[1](denoted {\bf I} henceforth).
 We give a brief review
of this relationship in Sec.2. In Sec.3, we present  the existence
proof of the fractal solution for a (time dependent) first order 
linear ODE .
\vskip 1cm
 
\section{Geometric phase and Time}
	In a time dependent quantal evolution the evolving state acquires 
a geometric phase apart from the usual dynamical phase. Let 
the evolution of the quantal state $\psi(t)$ be described by the 
Schrodinger equation

$${i\hbar {d \over{dt}} \psi =H \psi} \eqno (1)$$

\noindent where $H$ is a time dependent Hamiltonian operator and $t$ is
the external Newtonian time. As time flows, the quantal state $\psi$
($< \psi | \psi > =1$) traces a trajectory in the 
 Hilbert space $\cal H$, considered as a U(1) 
principal bundle over the projective space $\cal P$. Due to the irreducible
quantal uncertainty the actual path must always be fluctuating in nature
thus allowing the evolving state to follow a path intersecting 'near by'
rays in ${\cal P}$. If the state is assumed to move along a cyclic path
in ${\cal P}$ it returns eventually to the initial ray, but with a phase
difference. In the case of an open path, the phase difference between the 
initial and final states equals to that of the cyclic path obtained by 
joining the initial and the final rays in ${\cal P}$ by the shortest 
geodesic. The total phase $\gamma$ so acquired by the state now consists of 
two 
components: $\gamma = \gamma_d + \gamma_g$, the dominant 'dynamical' and 
the relatively small 'geometric'
phases respectively. The dynamical  phase 
$\gamma _d = \hbar ^{-1} {\int {h(t) dt}}$
where $h(t) = <\psi | H \psi >$ denotes the mean energy 
in the state, is a consequence of the mean dynamical evolution in the state. 
This mean evolution could  further be realized as a pure verticle 
displacement of the state along the the fibre of its ray in the associated
Hilbert bundle. Further, the scale of the external time $t$ is set by the mean 
energy : $t \sim {1 \over {h_0}}, h_0 = h(t_0),t_0$=the initial time.
As shown in {\bf I},the geometric phase ${\gamma_{g}=\int{A}}$ where 
$A= -i<\phi | d|\phi>$ and $\phi =e^{i\gamma} \psi$; on 
the otherhand could be interpreted as an effect of the inherent fluctuations
in the actual quantal evolution. Although, the nature of this phase is 
of course geometric, as encoded in the associated parallel transport law, 
it appears ,in any way, as a correction term in the mean energy, thus 
offering itself  to an equivalent dynamical (Hamiltonian) treatment.
The geometric phase thus provides naturally a new (microscopic) time 
scale in the evolution, having a conjugate relationship instead with 
the energy uncertainty, rather than with the mean energy. Following the 
  Leibnizian view[2], we call this time scale(variable) the intrinsic 
(geometric) time. Interestingly, this dynamical treatment of the 
geometric phase then allows one to write down a scaled Schrodinger equation  
which is suitable for an independent dynamical description of the fluctuating 
state. The reproduction of a(n) (almost) 
self-similar replica of the Schrodinger equation in the scale of a 
fluctuation involves a sort of a 'renormalization' in the original 
equation which amounts to a subtraction of the mean (dynamical) motion in 
the state. It also turns out (see below) that the intrinsic time scale 
 relates inversely (dually) to the extrinsic one in the limit of 
 large fluctuations O(1) (strong interaction) in the state.
Thanks to  the irreduciblity of the quantal uncertainty, one can  then 
iterate the above steps an unlimited number of times, thus  arriving at 
an interesting observation: {\it quantal fluctuations are essentially 
fractals:  having self-similar structures at all time scales}. As a corollary,
the trajectory of the evolving nonstationary state and hence the complex 
state function itself must also be fractals(at least in the time variable).
The later conclusion was not stated explicitly in {\bf I}. We shall 
supply a proof in favour of this in the next section.

  A remark is in order here. Quantum mechanics is a linear theory, with only 
a first order time derivative in the Scrodinger equation. Further, the 
(quantal) uncertainty
principle is a direct outcome of this inherent linearity. Any  linear 
time invariant model e.g., one in a signal processing system, also 
exhibits time-energy uncertainty. The above analysis should therefore be 
applicable not only in the domains stated above but also to any first order
 linear ODE with nontrivial time dependence. We recall that the 
(nonadiabatic) geometric phase was historically first discovered not in 
quantum mechanics but in optical polarisations having a linear
Schrodinger equation-like guiding equation.

\section{Main Results}

	Let us consider a first order linear ordinary differential 
equation of the form

$${i {d \over {dt}}y = h(t)y} \eqno{(2)}$$

\noindent where $h(t)$ is a real function  of the real parameter
$t$. The factor $i$ in the l.h.s. of eq(2)  is introduced to keep the analogy  
with the Schrodinger equation in view. This is not necessary,however, in 
general. The complex functions $y(t)$ are assumed to belong to a linear
(Hilbert) space.
The corresponding projective space thus consists of the rays
$y^{\prime}={e^{i\alpha}y,\alpha}$ real. For the sake of clarity, we assume 
eq(2) to represent a linear time invariant dynamical system  with mean 
'energy' $h_0=h(t_0)$ at the intial time $t=t_0$. Eq(2) is thus (form) 
invariant
under the group of translations, which in turn gaurantees the uniqueness of
the standard exponential solution $y=exp(-i\int_{t_0}^{t}h(t)dt)$. We however
wish to present  a new class of fractal solutions to eq(2) by extending the 
group of translations to the (modular) group SL(2,R). We remark that the 
existence of a {\it nontrivial} geometric phase naturally provides the 
necessary
window for introducing the duality transformation in a time dependent 
(linear) dynamical system. However, it turns out that the technique 
developed in {\bf I} on the basis of a geometric phase is more general
and could in fact be used in a situation with a vanishing geometric phase.

       Note that in a quantum mechanical model $h(t)$ stands    
for a time dependent Hamiltonian operator. In an ordinary classical situation 
eq(2) may either be considered as a matrix equation e.g., the Fermi-Walker 
transport equation for the polarised light or be assumed to involve a 
background 
source of a motive force, to generate a nontrivial
geometric phase. We however restrict the discussion  to the simplest case
where $y$ denotes a single complex function having continuous first 
derivative.We show that SL(2,R) acts as an invariance group even in this 
apparently trivial equation.
   
    Let us proceed  mimicking the steps of {\bf I}  thus indicating briefly 
how an intrinsic time variable $\tau$ appears in eq(2). Because of the 
time translation invariance, it is sufficient to study the nature 
of the solution close to the initial time which we now set to $t=0$.
Further, assume $t$ to be a dimensionless variable measured in the unit of
${1\over h_0}$.Thus the Taylor expansion of $h(t)$ close to 0 in eq(2) yields

$${i{d\over{dt}}y=\left[1+{\nu(t)}\right]y} \eqno(3)$$

The time dependent term in the rhs of this equation could be interpreted
as the correction  due to 'fluctuations' over the mean 'energy'. For a very 
short time interval (i.e.,neglictng $O(t^2)$ terms),fluctuations can be 
approximated 
as $\nu(t)={\nu_1}t$. In the external time scale fluctuations thus scale as $t$.
However, ${\nu}$ gives rise to a new time scale at the level of the first-order 
fluctuation, which follows from the equivalence of the description of the 
present model with that in {\bf I}. Let us denote this intrinsic time scale  
by ${\tau \sim {1 \over\nu_1}}$. To make the analogy with {\bf I}
complete let us write the 'Born-Oppenheimer' anstaz $y=y_0y_1, y_0=e^{-it}$,
and introduce the intrinsic time ${\tau}$ through

$${{d\over{dt}} -1  \equiv t{d\over{dt}}=- {d\over{d\tau}}} \eqno(4)$$

We thus obtain 

$${-i{d\over{d\tau}}y_1 = \nu_1\left[1+\nu_2(\tau)\right]y_1} \eqno(5)$$

Eq(5) has the following interpretation.The operator in the lhs of (4) 
subtracts out the mean (dynamical) evolution in (3) that manifests in the 
unit of the external time $t(\sim 1)$. The residual (renormalized) evolution 
now consists only of 
the small scale fluctuations. However, in the scale of the intrisic time 
${\tau \sim \nu{_1}^{-1}}$, an externally small fluctuation does appear 
substantially large. Eq(5) then tracks this 'internally large' fluctuation
exactly in the same spirit as that of the original eq(3): i.e.; by 
identifying ${\nu_1}$ as the (scaled) mean 'energy' with a relatively
small  correction ${\nu_{2}(t)}$ from the 
(2nd order) fluctuations. Further, the switching of the treatment from the 
external time scale down to the intrinsic scale amounts,in fact, to a 
transformation of the ordinary time $(t)$ scale to a logarithmic scale: 

$$\tau = ln|t|^{-1} \eqno(6)$$

This is made explicit in the second equality in eq(4). Eq(5) thus offers 
a new approach in probing the very short time evolution of eq(2). Note that 
the 
standard exponential solution is obtained from the initial 'mean' 
solution $y_0$  near  $t=0$ through a combination of two operations: i)by 
improving upon the initial solution with the inclusion of the neglected 
terms in the Taylor (perturbation) series expansion (the Picard's 
iteration), and then ii) by the 
successive applications of the translation group 'horizontally' along
the ordinary $t$-axis. The first operation could indeed be achieved through 
the subtraction proceedure (c.f.,the lhs operator in eq(4)), but continuing 
to work instead in the original time (scale) $t$. The new possibility 
that emerges 
in the present discussion is the following.Instead of following the evolution 
further 'horizontally', one could choose to dive  'vertically' down to the 
(intrinsic) logarithmic scale, thus climbing ,as if, to a {\it ripple} of 
the background fluctuation. Consequently, the scale of the fluctuations gets
sufficiently magnified (stretched), thus making a room for an (almost) 
independent  treatment of the same, analogous to the mean motion.
   
    To continue, we note that the (-) sign in eqs(4) 
and (5) is typically a consequence of the geometric origin of the intrinsic
time introduced in {\bf I}. It reveals a sort of 'relativity' between the 
extrinsic and the intrinsic treatments: the direction of traversal of a path
as seen from the external time frame gets reversed in the intrinsic scale.
Note also that the subtraction operation of the mean energy in every unit of
time (e.g., $t \sim 1$) amounts to a folding( squzeeing) 
on the solution space.
Eqs(3)-(5) thus encode a set of successive  stretching -twisting and folding 
operations; which can clearly be iterated ad infinitum; thus establishing,
in turn, a hierarchy of self-similar structures in the very short time 
limit of the evolution.
  A number of observations can now be made.

1.Eq(6) represents a generalized duality relation between t and ${\tau}$.
In the limit of small ${\tau}$, this reduces to $t\approx {\tau^{\prime}}^{-1},
\tau^{\prime} =1+\tau$. Thus in the limit of large fluctuations, the
SL(2,R) group is realized as the exact invariance group of the ODE (2).

2. The duality eq(6) also tells us that an exact replica of the original
eq(3) is reproduced after the  second iteration when $t\sim 0$. Succintly, 
the complete iteration process can be expressed as

$$
\begin{array}{ll}

i{d\over dt_n}y_n & = (-1)^n \nu_n \left[1 + \nu_{n+1}(t_n)\right]y_n\\ 
	    t_{n+1} & = -ln(\nu_{n}t_n), t_0\equiv t, \nu_0=1, n=0,1,2,\ldots 

\end{array} \eqno(7)
$$
In the limit $t_n \rightarrow 0$, the duality transformations reduce to 
the scaling relations $t_{n+1}=\nu_{n}t_n$. For, $t_n \rightarrow \nu_{n}^{-1}$
in the log scale $\Rightarrow t_{n+1}= \nu_n(\nu_{n}^{-1} - t_n)$ and 
then replacing $\nu_{n}^{-1} -t_n$ by $t_n$ (because of the 
translation invariance) one obtains the result. Thus the set of nonlinear
operations, as detailed above, defines a hyperbolic iterated function
system (IFS)[3] with scaling (contractivity) factor $\nu_1 
(= max(\nu_1,\nu_2)) , 0\leq \nu_1 <1$, at each point $ t=t_0 $ of the 
real t-axis. The intended fractal solution of eq(2) ( at $t= t_0$), our 
main result, is thus obtained as the unique attractor of this IFS. 

3. Interestingly, the limiting value of the scaling factor $\nu_n$, for 
any $n$, at the fixed point $t_n=0$ is given by the 'universal' value:
$\nu_{g}={{\sqrt5 -1}\over 2}$, the golden mean.This follows from the 
necessary constraint that the system of equations (7) must coincide at 
the fixed point. One thus obtains the relations $ \nu_{n+1} (1+\nu_{n+1})
=\nu_{n}^{-1} (1+\nu_{n+2}) =(\nu_{n}^{_-1}\nu_{n+1}^{-1})(\nu_{n+1}^{-1}
\nu_{n+2}^{-1}) \ldots =1$, by virtue of the duality $t_n \sim \nu_{n}^{-1}$
but $\nu_{n+1} \sim t_n$ ($\Rightarrow -{\nu_{n+1}^2{dt_{n+1}\over dt_n}}
\sim 1$).
 
4. As a consequencec,the limiting form of the set of iterated equations    
(7), at the fixed point, is given by 

$$
\begin{array}{ll}
i(d /{dt_n})y_n  &= (-1)^n \nu_g y_n \\ 
t_1&=t , t_{n+1}=\nu_{g}t_{n}, n> 0  
\end{array} \eqno (8) $$

\noindent These equations could be interpreted as one obtained by splitting the 
infinite degenarecies of the original equation $i{dy \over dt}=y$, through
the repeated applications of the nonlinear stretching and  twisted -folding
transformations at the fixed point. Indeed, this could be realized by 
introducing {\it a} partition of the unity: $1=(1-\nu_n) +\nu_{n}$,
identifying the braketed term as the 'mean' and the remaider as 
corrction due to the fluctuations
and then following the above steps. The apparent arbitrariness in the 
partition gets washed away at the fixed point, thus leading to the 
unique limiting system ,eq(8).One thus obtains eq(8) as the attractor 
of the class of ODE considered here, under the nonlinear invariance group  
SL(2,R). An explicit form of the fixed point fractal solution could 
therefore be constructed  in the form  
 
$$ {y=exp(-i\left[t-\nu_{g}(t_1-t_2 +t_3 -\ldots )\right])} \eqno(9)$$

\noindent where the parameters $t, t_n$ are treated as independent variables
and each tends to 0 satisfying $t\geq t_1 >t_2 > t_3> \ldots $.
By  duality the point $t_n=\nu_{n}^{-1}$ is mapped to $t_{n+1}=0$,for each n,
ensuring a nontrivial sewing of the ordinary exponentials along the 
 (internal)  'verticle' direction. However,the function (9)
collapses to the simple form $y\sim exp(-i\nu_{g}t)$ in the very short 
time limit, provided one makes use of the scaling relations $t_1=t, 
t_{n+1}=\nu_{g}t_n $. 

  To interprete this result, we note that the original equation 
$dy/d{\tilde t} =y, \tilde t = -it$, in the log scale 
$|\tilde t|=ln{\tau}^{-1}$, can be translated as the 
definition of the box-counting dimension of the real axis parametrized by
$\tilde t$ : $ \lim \limits_{\tau \rightarrow 0}{lny \over ln{\tau^{-1}}} =1$, 
provided $y$ is identified as the total number of infinitesimal intervals
 needed to cover a finite segment of the $t$-axis.. In the present
case, writing $\nu_g =1-d, d={{3-\sqrt 5} \over 2}$, we get instead a 
nontrivial scaling law $y \approx \tau^{-(1-d)}$. Following {\bf I}
we interprete this result as one revealing a fractal structure in the
real $t$ axis itself. The real $t$ axis thus behaves as a fat fractal with 
{\it exterior dimension} $d$[4]. The golden mean is thus realized as the 
corresponding {\it uncertainty exponent} of this fat  fractal[1,4] . 
 
4.The above discussions also clearly reveal the presence of the 
deterministic chaos in the very short time evolution of a linear 
dynamical system.

\section{Final Remarks}
 
 We have presented a new  approach in analyzing a (linear) differential
equation. The results discussed here are expected to find interesting
applications in a number of physical problems where a perturbative 
approach normally fails. Nevertheless, it's striking  how the 
golden mean emerges as a universal scaling constant in probing
the short distance (time) structure of a linear dynamical system.

\end{document}